\documentclass[conference]{IEEEtran}
\IEEEoverridecommandlockouts
\usepackage{cite}
\usepackage{amsmath,amssymb,amsfonts}
\usepackage[ruled,linesnumbered]{algorithm2e}
\usepackage{subcaption}
\usepackage{booktabs}
\usepackage{multirow}
\usepackage{graphicx}
\usepackage{textcomp}
\usepackage{enumitem}
\usepackage{xcolor}
\usepackage{listings}
\usepackage[hidelinks]{hyperref}
\def\BibTeX{{\rm B\kern-.05em{\sc i\kern-.025em b}\kern-.08em
    T\kern-.1667em\lower.7ex\hbox{E}\kern-.125emX}}
\lstdefinestyle{mystyle}{
	commentstyle=\color{codegreen},
	keywordstyle=\color{magenta},
	numberstyle=\tiny,
	stringstyle=\color{codepurple},
	basicstyle=\scriptsize\ttfamily,
	breakatwhitespace=false,         
	breaklines=true,                 
	captionpos=b,                    
	keepspaces=true,                 
	numbers=left,                    
	numbersep=5pt,                  
	showspaces=false,                
	showstringspaces=false,
	showtabs=false,                  
	tabsize=2,
	aboveskip=0pt,
	belowskip=0pt,
	xleftmargin=2em,
	xrightmargin=0em,
	frame=topline,
	mathescape=true,
	fontadjust=true
}

\begin{document}

\title{PRAGMA: A Profiling-Reasoned Multi-Agent Framework for Automatic Kernel Optimization}

% \and
% \and
% \and
% \and
% \and
\author{
	\IEEEauthorblockN{
		Kelun~Lei,
		Hailong~Yang,
		Huaitao~Zhang, Xin~You, Kaige~Zhang,  Zhongzhi~Luan,
		Yi~Liu, and~Depei~Qian
	}
	\IEEEauthorblockA{
		School of Computer Science and Engineering, Beihang University, Beijing, China\\
		\{kelunlei, hailong.yang, zhanght329, youxin2015@buaa.edu.cn, kaige.zhang@buaa.edu.cn, 07680, yi.liu, depeiq\}@buaa.edu.cn\\
	}
}

\maketitle

\begin{abstract}
Designing high-performance kernels requires expert-level tuning and a deep understanding of hardware characteristics. Recent advances in large language models (LLMs) have enabled automated kernel generation, yet most existing systems rely solely on correctness or execution time feedback, lacking the ability to reason about low-level performance bottlenecks. In this paper, we introduce PRAGMA, a profile-guided AI kernel generation framework that integrates execution feedback and fine-grained hardware profiling into the reasoning loop. PRAGMA enables LLMs to identify performance bottlenecks, preserve historical best versions, and iteratively refine code quality. We evaluate PRAGMA on KernelBench, covering GPU and CPU backends. Results show that PRAGMA consistently outperforms baseline N-PRAGMA without profiling enabled and achieves 2.81$\times$ and 2.30$\times$ averaged speedups against Torch on CPU and GPU platforms, respectively.
\end{abstract}

% \begin{IEEEkeywords}
% component, formatting, style, styling, insert
% \end{IEEEkeywords}

\section{Introduction}
\label{sec:intro}
% the emergence of AIKG
Optimizing computational kernels is fundamental to achieving high performance in modern AI and HPC systems. Traditionally, reaching near-peak efficiency has required extensive manual tuning and deep expertise in architecture-specific optimization, making the development and maintenance of high-performance kernels both labor-intensive and error-prone. Recently, the rapid emergence of large language models (LLMs) has ushered in a new era of AI-assisted code generation, enabling models to automatically synthesize domain-specific compute kernels such as matrix multiplication, convolution, and reduction operators. These advances hold the promise of significantly accelerating the development of optimized kernels across diverse hardware platforms.

% 传统调优方法的局限性
% 复杂精细的手工算子调优可以实现较高的硬件性能，但是
% 补充一些rl相关的（需要训练）-> 无需微调的LLM工作
Manual kernel optimization can achieve high performance through fine-grained tuning. However, the rapid evolution of hardware and proliferative ML architectures~\cite{tay2022efficient} often invalidates previous optimizations~\cite{dao2022flashattention,dao2023flashattention,shah2024flashattention}, requiring repeated manual effort. Recently, LLM-based kernel generation~\cite{ouyang2025kernelbench,rahman2025marco,wei2025astra,wen2025multikernelbench,zhou2025qimeng,zhang2025qimeng,zhou2025qimenga,aikg2025git} has emerged as a promising alternative. These systems can automatically generate code and iteratively refine it based on compilation and execution feedback. For instance, AIKG~\cite{aikg2025git} leverages iteratively orchestrated multiple LLM agents for effective kernel generation. Yet, most existing methods focus on generic, hardware-independent optimization. Their feedback is often restricted to correctness or overall runtime. Such coarse-grained feedback provides limited guidance for performance reasoning and often leads to unstable results. Only a few recent studies attempt to incorporate fine-grained performance metrics. For instance, SwizzlePerf~\cite{tschand2025swizzleperf} leverages L2 cache miss information to guide memory access pattern optimization for AMD MI300x~\cite{smith202411} disaggregated GPU, yet its profiling scope is narrow. Consequently, enabling LLMs to reason over detailed architectural characteristics and systematically utilize profiling feedback to guide optimization remains an open and underexplored challenge.
In this work, we present PRAGMA, a performance-guided multi-agent system for high-performance kernel generation.
Unlike prior LLM-based approaches that rely solely on execution time or correctness feedback, PRAGMA introduces a profile-guided multi-agent architecture that establishes a closed feedback loop between code generation and performance evaluation.
To provide detailed and hardware-aware feedback, PRAGMA employs Profiler Agent which gathers low-level metrics with its semantic information from diverse profilers such as Nsight Compute and Linux perf across CPU and GPU platforms.
Based on these data, PRAGMA leverages a dedicated Conductor Agent to interpret profiling results and perform bottleneck classification, mapping low-level performance metrics to high-level optimization hints. Guided by these insights, the Coder Agent iteratively generates and refines the kernel implementation.
Note that each iteration preserves the historically best kernel implementation and its corresponding profiling data, enabling the LLM agent to reason about performance bottlenecks in context.
We evaluate PRAGMA on both GPU and CPU platforms across KernelBench~\cite{ouyang2025kernelbench}, which is a comprehensive benchmark suite of machine learning kernels. Experimental results show that PRAGMA consistently outperforms prior LLM-based systems, achieving a speedup of 2.81$\times$ compared to Torch on CPU and speedups of 2.30$\times$ and 4.50$\times$ over Torch and Caesar~\cite{ouyang2025kernelbench} baselines on GPU. Therefore, PRAGMA can provide interpretable, profiler-driven reasoning that bridges human performance engineering and automated code generation.

% contributions，进一步精简profiler的贡献，仅需提及关键的模块化可扩展接入新profiler，profiler知识库自动构建，以及自适应的profiler指标采集。
Our main contributions are as follows:
\begin{itemize}[leftmargin=*,align=left]
	\item We propose a performance-guided multi-agent framework PRAGMA that integrates rich profiling feedback into LLM-based kernel optimization.
	\item We design and implement a cross-platform profiling submodule supporting both GPU and CPU backends with unified performance data extraction.
	\item We propose a bottleneck-aware reasoning module that associates performance metrics with concrete optimization strategies, enabling interpretable and architecture-specific kernel improvement through iterative and systematic generation.
	\item We evaluate PRAGMA across KernelBench, demonstrating up to 10.95$\times$ performance speedup over baseline LLM-generated kernels. % over pytorch?
\end{itemize}
\section{Background and Motivation}
\label{sec:background}
% kernelbench, swizzleperf, astra, AlphaCode, Tritonbench, Alphaevolve, Llm-vectorizer, Vectran,Improving parallel program performance with LLM optimizers via agent-system interfaces,Agentcoder: Multi-agent-based code generation with iterative testing and optimisation,Metagpt: Meta programming for a multi-agent collaborative framework, Trace is the next autodiff: Generative optimization with rich feedback, execution traces, and llms, Autogen: Enabling next-gen llm applications via multi-agent conversations, Cuda-l1: Improving cuda optimization via contrastive reinforcement learning, “Cuda-llm: Llms can write efficient cuda kernels

% 1. 有很多评估LLM写代码（cuda or dsl）的能力与性能的benchmark，结论是，挑战是。。。
% 2. 有很多单或者多agent做向量化、cuda kernel、强化学习、反馈指导等的工作了，但它们。。。

% 单个agent负责这么多事情，需要了解的东西太多，prompt过长，效果不好
% 不加profiling信息，LLM不知道优化方向，不知道代码优劣在哪里，只能做general的优化
% 不加错误反馈，不好修复bug，可能LLM就重复生成相同的kernel或者直接重写，无用功很多
% single agent work
Recent studies on program code generation with LLMs have shown a great number of impressive results.
Evaluations~\cite{liu2023your,nguyen2022empirical} on various benchmarks demonstrate the superior coding ability of LLMs considering different aspects such as correctness, efficiency, and maintainability. % qimeng-gemm
% GPT-4 achieves a 76.2\% pass@1 rate on EvalPlus and HumanEval+~\cite{liu2023your}. DeepMind’s AlphaCode reached the top 54.3\% among human competitors in online programming contests, and GitHub Copilot demonstrated particularly strong results in Java, solving 57\% of LeetCode problems~\cite{nguyen2022empirical}.
% 单个agent负责这么多事情，需要了解的东西太多，prompt过长，效果不好
Despite these advances, most single-agent systems still exhibit limitations.
A single LLM must handle problem understanding, algorithm design, code synthesis, verification, and optimization simultaneously. This requires very long prompts and complex in-context reasoning, which often exceed the model’s effective context window.
% 多agent可以解决这个问题，现有多agent的工作有：
To address these limitations, researchers have proposed multi-agent frameworks~\cite{wu2024autogen,hong2023metagpt,qian2023chatdev} that decompose programming tasks into specialized roles coordinated through dialogue and reasoning.
In particular, many frameworks~\cite{ouyang2025kernelbench,wen2025multikernelbench, rahman2025marco,wei2025astra,tschand2025swizzleperf} have further explored the ability of LLMs to write GPU kernels.
These systems enable collaboration between agents responsible for planning, coding, testing, timing, and refinement. 
% 然而HPC challenges，无论是单agent还是多agent，都无法很好的解决
However, HPC programs demand a deep understanding of architectural and algorithmic principles such as cache locality, memory hierarchy, vectorization, and multi-threading. General-purpose LLMs, trained predominantly on serial, high-level programming data from public repositories, lack exposure to domain-specific optimization patterns. Consequently, existing works cannot effectively connect code generation with quantitative performance feedback, causing iterative refinements to yield unpredictable or inconsistent performance outcomes.
This motivates our performance-guided multi-agent system tailored for HPC code generation.
\section{Methodology and Implementation}
\label{sec:method}
% 实验可以用8 irregular matmul来写一个case study，分析profiling信息和历史最优代码的作用，给出conductor和coder的reasoning
% overview图
\begin{figure*}[htbp]
	\centering
	\includegraphics[width=\linewidth]{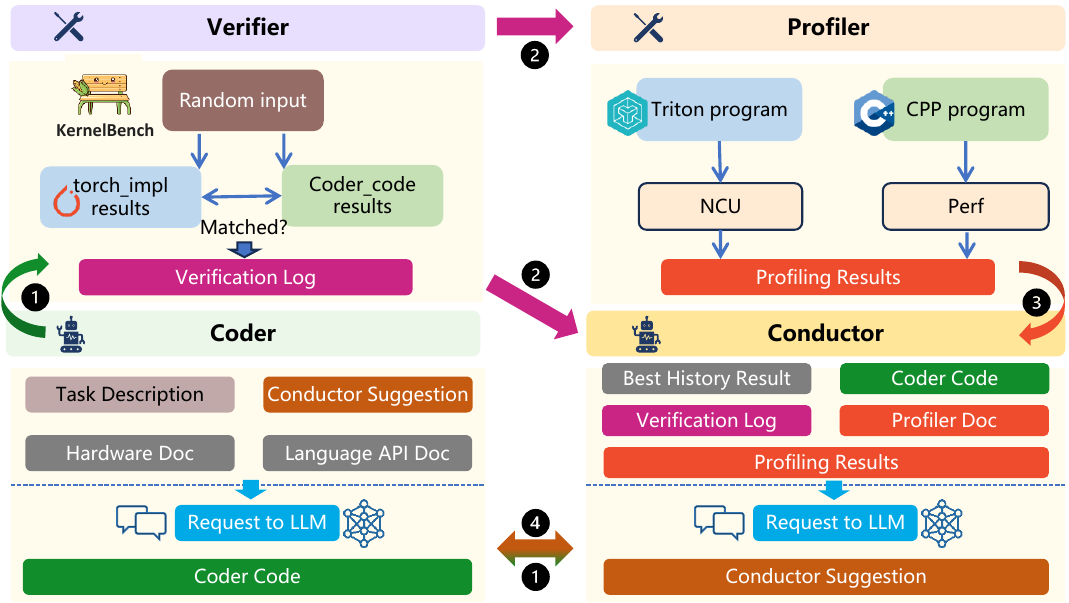}
	\caption{Overview of PRAGMA. The code generation and optimization process shows the interaction between the Coder, Verifier, Profiler, and Conductor agents. The system begins with an input task description to Coder.}
	\label{fig:overview}
\end{figure*}

\subsection{System Overview of PRAGMA}
\label{sec:method:overview}
% 分别介绍几个agent的角色
% conductor: conductor负责收集其他Agent的结果，包括Coder生成的代码，verifier的验证结果，包括可能存在的错误信息，以及profiler采集的profiling信息。在此基础上，conductor还会基于这些信息给其他agent的任务提供建议。conductor的prompt组成包括编译错误以及运行时错误的可能原因与解决办法，profiler说明文档、来自verifier的error log以及来自profiler agent的profling result等等。
At the core of the proposed system lies a multi-agent architecture designed to achieve performance-guided kernel generation and optimization. Each agent in the system is assigned a distinct role and collaborates through structured information exchange under the coordination of a central Conductor Agent. Together, these agents form an adaptive, feedback-driven loop that progressively improves both the correctness and efficiency of automatically generated kernels.

\textbf{\textit{Conductor Agent}}: The Conductor Agent serves as the central orchestrator of the system, responsible for aggregating and interpreting the outputs from all other agents. It collects the source code produced by the Coder Agent, the verification feedback from the Verifier Agent, and the performance profiling data gathered by the Profiler Agent. Based on this comprehensive information, the Conductor Agent synthesizes contextual insights and formulates actionable guidance for subsequent iterations.

The Conductor's prompt is composed of several structured components, including:
(1) generated source code from the Coder;
(2) potential causes and remedies for compilation or runtime errors derived from Verifier logs;
(3) explanatory documentation from the Profiler describing the meaning and interpretation of performance metrics;
(4) detailed profiling results that capture hardware-level behavior and bottlenecks; and
(5) historically best-performing code with profiling results.

By integrating these heterogeneous inputs, the Conductor Agent acts as a reasoning and decision-making hub, enabling informed instructions to the Coder and other agents. 

\textbf{\textit{Coder Agent}}: The Coder Agent is responsible for code synthesis and refinement. Guided by the Conductor’s instructions, it generates kernel implementations that meet the desired functional and performance goals. The Coder incorporates diagnostic feedback, such as compiler errors, profiling reports, or performance optimization hints provided by the Conductor to iteratively improve the generated code. The Coder’s capabilities include both initial code generation from high-level specifications and incremental modification based on feedback, enabling a progressive refinement process similar to that of an expert human developer.

\textbf{\textit{Verifier Agent}}: The Verifier Agent ensures the functional correctness and executability of the generated code. After each iteration, it compiles and runs the Coder’s output, compares execution results against reference outputs, and records any compilation or runtime errors. As verification processes do not need LLM capabilities, we do not use LLM for Verifier.

\textbf{\textit{Profiler Agent}}: The Profiler Agent is responsible for performance data collection and interpretation. It executes the verified kernel on the target hardware and gathers low-level hardware performance metrics. 
% The Profiler Agent references an offline documentation base that describes the meaning and implications of each performance metric, enabling consistent and interpretable feedback. 
The collected data are then structured and provided to the Conductor Agent for identifying bottlenecks and suggesting targeted optimization strategies.

Through this tightly coupled interaction between the Coder, Verifier, Profiler, and Conductor Agents, the system forms a closed-loop optimization workflow that iteratively converges toward functionally correct, performance-optimized kernel implementations across heterogeneous hardware platforms.

\subsection{Design of Profiler Agent}
\label{sec:method:profile}
The Profiler Agent employs NVIDIA Nsight Compute (NCU) on GPU backend and Linux Perf on CPU backend to collect profiling data.
To ensure the LLM understanding of profiling data and provide a scalable and extensible approach to integrating diverse profilers to PRAGMA, we develop an automated profiling knowledge consolidation module that generates a unified offline compendium of profiling tool features, metric semantics, and diagnosable performance issues. 

The consolidation process proceeds in two stages.
In the first stage, raw documentation is automatically crawled from official online sources of major profilers, such as NVIDIA Nsight Compute for GPU backend and Linux Perf for CPU backend. Each raw text segment is independently summarized by an LLM to extract essential content, such as the functional description of profiling metrics, their underlying measurement mechanisms, and the performance bottlenecks they are designed to reveal. 

In the second stage, these intermediate summaries are fed into a higher-level synthesis process, where the LLM merges and refines the content into a comprehensive and coherent offline document. The resulting corpus standardizes terminology and organizes information across profilers into a common schema that supports both semantic search and performance reasoning.

% 补充配合LLM选取反馈指标的内容，指标太多，忽略不重要的指标，选取人类性能工程师常用的指标（有指向性）。
Additionally, profiling tools produce a large number of metrics, many of which are not critical for guiding performance optimization. To address this, we select a subset of metrics that are commonly used by human performance engineers and have direct diagnostic relevance. For GPU workloads, these include kernel execution time, occupancy, thread register usage, shared memory utilization, tensor core active cycles, and memory throughput. For CPU workloads, we focus on metrics derived from top-down microarchitecture analysis~\cite{yasin2014top}, such as retired instructions, front-end bound, back-end bound, bad speculation, and overall cycle counts. 
Furthermore, a key feature is that the Conductor Agent can dynamically specify the desired extra metrics. During each profiling run, these requested metrics are merged with the default set, collected, and incorporated into the feedback pipeline. 
By filtering for these key indicators, the Conductor Agent can provide concise, actionable feedback, highlighting critical bottlenecks, and guiding the next round of optimization. 

In sum, Profiler Agent can provide profiling knowledge and adaptive performance metrics, which not only enhances the Conductor Agent's ability to interpret performance data but also provides the Coder Agent with explainable feedback regarding kernel inefficiencies and optimization opportunities. 
Additionally, incorporating a new profiler requires only the addition of its documentation URL and the implementation of the profile collection APIs within PRAGMA, which has high extendibility and portability.
% This design effectively bridges the gap between low-level performance metrics and high-level AI-driven code optimization.

\subsection{Profile-guided Iterative Code Refinement}
\label{sec:method:iterate}
% 描述LOOP过程，信息采集与记录、接受、处理
% LLM无法一轮即可生成性能优秀的代码，甚至还可能出现生成代码无法编译、运行出错或者运行结果不对的问题。因此我们的系统支持迭代式的代码生成、修复与优化。具体而言，每轮Coder生成的代码都会首先交由verifier进行编译与运行以及结果比对，确保代码的正确性。此时，如果正确性校验未通过，错误信息将会提供给conductor agent并将错误修复建议输入到Coder Agent进行新一轮的代码生成。当正确性校验通过后，系统将会调用profiler对代码进行性能数据采集。系统之后对性能数据进行提取与结构化并填入conductor的prompt中。接下来，conductor会给出性能优化建议，指导Coder Agent针对性进行优化代码生成。
\begin{figure}[ht]
	\centering
	\includegraphics[width=\linewidth]{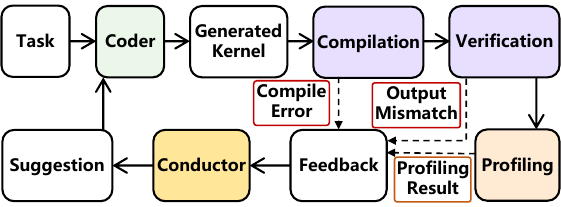}
	\caption{Profile-guided iterative optimization process. Coder agent leverages the suggestion provided by Conductor agent who receives the feedback from Verifier and Profiler to refine the code iteratively.}
	\label{fig:iter}
\end{figure}

High-performance code often cannot be generated by LLMs in a single attempt. In frequent cases, the generated code may fail to compile, produce runtime errors, or yield incorrect results. To address these issues, our system adopts an iterative paradigm of code generation, verification, and optimization as shown in Figure~\ref{fig:iter}.

Specifically, in each iteration, the Coder Agent first generates a candidate implementation, which is then passed to the Verifier for compilation, execution, and correctness checking. The Verifier ensures functional correctness by comparing the output of the generated code against reference results. If the correctness validation fails, the error logs and diagnostic information are forwarded to the Conductor Agent, which synthesizes targeted repair suggestions and provides them to the Coder Agent for the next round of generation.

Once the correctness check passes, the system invokes the Profiler to collect detailed performance data. These profiling results are then parsed, structured, and embedded into the Conductor’s prompt, allowing the system to reason about performance characteristics and bottlenecks. The Conductor Agent subsequently formulates performance optimization guidance, which directs the Coder Agent to perform targeted code refinements in the following iteration.
% 历史最优保留
To facilitate performance-aware reasoning across iterations, PRAGMA maintains a persistent record of the historically best-performing code and its corresponding profiling data. During subsequent iterations, the Conductor compares the current candidate’s profiling metrics with those of the best historical version to evaluate improvement or degradation. This comparison allows the system to reason about why performance changes occurred and to generate optimization instructions grounded in measurable performance behavior.
% 结束条件

Through this closed-loop iterative process, the system progressively enhances both the correctness and performance of the generated code, enabling LLMs to evolve from producing merely functional implementations to highly optimized, hardware-aware kernels.

\subsection{Multi-Backend Support}
\label{sec:method:multi-backend}
% triton-cuda, x86, arm?
% CPP + pybind
% 适配到新后端，仅需提供后端信息（cpu可以直接lscpu获得，加速器需要用户提供），以及后端profiler的文档。对于代码生成，triton支持很多加速器，如nv和amd的gpu，huawei ascend。c语言也可以在不同cpu上直接编译运行，调用自带的编译器即可。
Modern heterogeneous systems encompass a wide spectrum of hardware backends, including CPUs, GPUs from multiple vendors, and emerging AI accelerators. To enable portability and extensibility, PRAGMA is designed with a multi-backend architecture that allows seamless adaptation to new hardware platforms with minimal configuration effort.

Adapting PRAGMA to a new backend requires only two types of information:
(1) hardware specification data, and
(2) profiler documentation describing available performance metrics and their semantics.
For CPUs, key parameters such as core count and cache hierarchy are automatically extracted using tools like \verb|lscpu|.
For GPUs and accelerators, users provide basic specifications such as the number of compute units, memory bandwidth, and cache hierarchy.
For code generation on GPUs, PRAGMA adopts Triton~\cite{tillet2019triton} as a unified programming interface.
Triton also supports multiple accelerator backends, including NVIDIA GPUs, AMD GPUs, and Huawei Ascend NPU, enabling cross-platform kernel generation with minimal modification.
% During each iteration, the Coder Agent produces Triton kernels that can be directly compiled and executed across these accelerators.
For CPU targets, PRAGMA generates standard C++ implementations, which are compiled and executed through the system's native compiler via PyBind integration.
% For code generation, AIKG relies on Triton as a unified programming layer.
% Triton supports a wide range of accelerators, including NVIDIA GPU, AMD GPU, and Huawei Ascend.
% The Coder Agent generates Triton code that can be compiled and executed across these backends without modification.
% For CPUs, AIKG generates standard C++ code and compiles it with the system’s native compiler.
Additionally, PRAGMA provides a modular profiler interface.
Only the corresponding data collection interface needs to be implemented for each platform-specific profiler integrated into the Profiler Agent.
\section{Evaluation}
\label{sec:evaluation}

\subsection{Evaluation Setup}

\textbf{Environments.} % deepseek R1
All experiments are conducted on an Intel Xeon Gold 6230R CPU and a NVIDIA A100 40GB GPU. 
The CPU experiments use \texttt{gcc 12.3.0}, while the GPU experiments use CUDA \texttt{13.0} and driver version \texttt{580.95.05}. Meanwhile, the compilation flags are generated by the LLMs.

\textbf{Baselines.}
We compare our proposed PRAGMA framework with the following baselines:
\begin{itemize}
	\item (1) The reference Torch implementations.
	\item (2) N-PRAGMA i.e., PRAGMA w/o performance feedback.
  \item (3) Caesar, which adopts a single agent and is used in the multi-turn experiments in KernelBench~\cite{ouyang2025kernelbench}.
\end{itemize}
Additionally, we use DeepSeek-R1~\cite{guo2025deepseek} as the base LLMs for all the methods.
% These baselines represent, respectively, human-engineered kernels and AI-generated kernels without architecture-aware refinement.

\textbf{Datasets.}
We evaluate on KernelBench, a recently proposed benchmark primarily designed for assessing LLM-based GPU kernel optimization. We also use it to evaluate the effectiveness for CPU kernel optimization.
This benchmark contains diverse operator types including matrix multiplication, convolution, element-wise, and reduction kernels, covering various tensor shapes and computation patterns. 
For each task, we let all methods have a maximum of 15 attempts.

\textbf{Metrics.}
We report both \textit{Speedup} over the Torch baseline and generation \textit{Success} rate. 
Speedup is defined as the ratio of execution time between Torch and the generated kernel:
\[
\text{Speedup} = \frac{T_{\text{Torch}}}{T_{\text{Generated}}}.
\]
The \textit{Success} rate measures the fraction of kernels that compile, execute correctly, and pass verification.
In particular, to ensure stable measurement, runtime is measured using Triton do\_bench with 5 warmup runs and 100 repetition runs.

% PRAGMA with N-PRAGMA
\subsection{Performance: Speedup Distribution}
\label{sec:evaluation:overall}
\begin{figure*}[htbp]
	\centering
	\includegraphics[width=\linewidth]{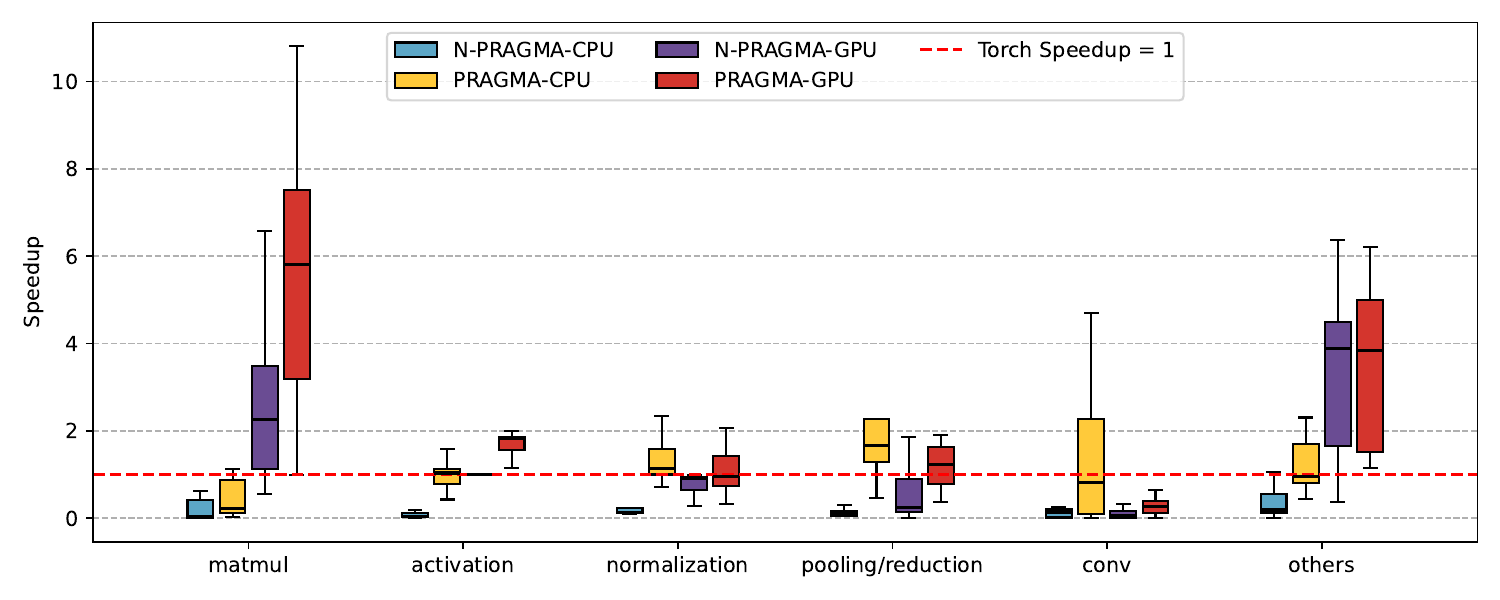}
	\caption{Performance of kernels from six KernelBench categories generated by PRAGMA and N-PRAGMA on CPU and GPU, with the reported speedup normalized to the Torch baseline.}
	\label{fig:overall_speedup}
\end{figure*}
% speedup = 1改成torch
% Figure~\ref{fig:overall_speedup}展示了PRAGMA、AIKG和torch在kernelbench上的总体性能box plot。我们将kernelbench的kernel归为6类，包括matmul、activation、normalization、pooling/reduction、conv和others，其中others包括一些loss和cumulative sum的kernels。
% Figure~\ref{fig:overall_speedup} summarizes the overall performance comparison across all kernels. 
% On average, PRAGMA achieves a 2.81$\times$ speedup over Torch and a 10.95$\times$ improvement over N-PRAGMA on CPU.
% For GPU kernels, the average speedup of PRAGMA reaches 3.74$\times$ compared to Torch and a 1.90$\times$ improvement over N-PRAGMA.
% 可以看到，在GPU上，PRAGMA和AIKG生成kernel的平均性能均优于Torch。PRAGMA由于引入了profiling信息，获得的加速比更为显著。而对于CPU，profiling信息的引入作用更加明显。
% This result validates the effectiveness of our performance feedback and iterative optimization mechanisms.

% PRAGMA在matmul上实现了最优加速， gpu好因为triton提供了autotune，每轮gpu都会根据性能数据更新autotune的候选参数，因此性能稳定上升。cpu上没有autotune，因此很难找到综合考虑向量化、并行、cache分层以及预取的tiling参数。尤其kernelbench中存在不同shape的矩阵kernel，LLMs难以在所有shape下都实现较优的性能。
% 但是，从图中还可以看到，无论是否引入profiling信息，在卷积类别的kernel中性能均没有表现出相较于torch的明显优势，甚至相差很多。这是因为卷积类的kernel相较于其他类别实现更为复杂，涉及循环嵌套层次更深，且输入张量的维度增加，做tiling、内存布局改善，prefetch等优化出错的概率增加，需要多次尝试才能保证正确性，因此性能没有明显优势。
Figure~\ref{fig:overall_speedup} presents the speedup distribution of PRAGMA and N-PRAGMA over Torch baseline across the \textsc{KernelBench} dataset.
The benchmark kernels are grouped into six categories: \emph{matrix multiplication (matmul)}, \emph{activation}, \emph{normalization}, \emph{pooling/reduction}, \emph{convolution (conv)}, and \emph{others} (including loss and cumulative-sum kernels).

On the CPU, PRAGMA achieves an average of 2.81$\times$ speedup over Torch and 10.95$\times$ improvement over N-PRAGMA.
On the GPU, PRAGMA delivers a 2.30$\times$ speedup over Torch and a 1.90$\times$ improvement compared with N-PRAGMA.
Meanwhile, it can be observed that both PRAGMA and N-PRAGMA generate GPU kernels that outperform Torch on average.
The integration of detailed profiling information allows PRAGMA to achieve a more significant improvement.
On the CPU, this effect becomes even more pronounced, as profiling data provides richer insights into memory access patterns and cache utilization which have a greater impact on CPU workloads.
These results confirm the effectiveness of our profile-guided feedback loop and iterative optimization mechanism, which enable the system to progressively refine performance.

Additionally, PRAGMA achieves the highest speedup on matrix multiplication kernels on GPU. This is achieved by adopting the Triton's autotuning mechanism. Coder can generate \texttt{triton.autotune} decorator for the Triton kernel and update the candidate parameters based on the profiling data in each iteration, ensuring steady performance improvement.
In contrast, using C++ on CPU backend lacks such autotuning support.
Finding optimal tiling configurations that balance vectorization, parallelism, cache hierarchy, and prefetching becomes challenging.
Moreover, KernelBench contains matrix kernels with diverse shapes, making it difficult for LLMs to achieve better performance across all cases compared to the BLAS library used by Torch.

Although PRAGMA consistently outperforms Torch and N-PRAGMA across most categories, we observe limited gains in convolution kernels.
Convolution operators exhibit substantially higher implementation complexity, involving deeper loop nests and higher-dimensional tensor accesses.
This complexity increases the risk of suboptimal tiling, layout transformation, and prefetching decisions during code generation.
As a result, achieving correctness and performance simultaneously often requires multiple iterations, leading to less pronounced acceleration in this category.

% success失败的都是conv，印证前面
\subsection{Comparison with Caesar}
\label{sec:evaluation:kernelbench}
% \begin{table}[t]
% \small
% \centering
% \caption{Overall performance comparison of LLM Agents on KernelBench. Speedup is normalized to the Torch baseline, and Fast$_1$ measures the fraction of kernels matching or exceeding Torch performance.}
% \label{tab:overall_performance}
% \renewcommand{\arraystretch}{1.2}
% \setlength{\tabcolsep}{6pt}
% \begin{tabular}{l|ccc}
% \hline
% \textbf{Agent} & \textbf{Success (\%)} & \textbf{Speedup ($\times$)} & \textbf{Fast$_1$ (\%)} \\ 
% \hline
% KernelBench & 100.0 & 1.00 & -- \\
% N-PRAGMA & 82.3 & 1.41 & 56.2 \\
% PRAGMA & 94.5 & 2.73 & 79.8 \\
% \hline
% \end{tabular}
% \end{table}

\begin{table}[t]
\small
\centering
\caption{Overall performance comparison of LLM Agents on the KernelBench Tasks. Fast$_1$ represents the percentage of problems for which the agent can generate custom kernels that are correct and as fast as the Torch baseline. Speedup is normalized to the Torch baseline.} % 解释fast1
\label{tab:overall_performance}
\renewcommand{\arraystretch}{1.2}
\begin{tabular}{l|c|cc}
\hline
\multirow{2}{*}{} & 
\multirow{2}{*}{\textbf{Success (\%) $\uparrow$}} & 
\multicolumn{2}{c}{\textbf{Torch}} \\ 
\cline{3-4}
 & & \textbf{Speedup ($\times$) $\uparrow$} & \textbf{Fast$_1$ (\%) $\uparrow$} \\ 
\hline
Caesar & 49.0 & 0.51 & 14.3 \\
N-PRAGMA & 92.0 & 1.28 & 32.2 \\
PRAGMA & 92.0 & 2.30 & 59.8 \\
\hline
\end{tabular}
\end{table}

We further compare PRAGMA against the baseline Caesar~\cite{ouyang2025kernelbench} on GPU platform since Caesar only supports CUDA kernels generation. 
Caesar relies on an agent to generate CUDA kernels and integrates them into PyTorch via PyBind. 
However, its feedback loop is limited to correctness checking and coarse-grained runtime measurement at the PyTorch layer. 
As a result, the system often exhibits high error rates and unstable performance, since the agent lacks visibility into hardware-level behaviors such as occupancy, resource utilization, memory throughput, and instruction-level inefficiencies.

Table~\ref{tab:overall_performance} summarizes the overall results. 
Caesar achieves a 49.0\% \textit{Success} rate and its generated kernels frequently fail to outperform the original Torch implementations. 
N-PRAGMA improves upon this by introducing multi-agent architecture, achieving 92.0\% \textit{Success} rate, 1.28$\times$ average speedup, and 32.2\% Fast$_1$. 
Nevertheless, without profiling-guided insights, N-PRAGMA’s optimization remains largely heuristic and insensitive to architectural bottlenecks.

In contrast, PRAGMA introduces a performance-guided multi-agent framework that integrates profiling data from Nsight Compute. 
This enables fine-grained bottleneck classification and targeted optimization reasoning. 
As shown in Table~\ref{tab:overall_performance}, PRAGMA achieves 2.30$\times$ average speedup with 59.8\% Fast$_1$, consistently outperforming both Caesar and N-PRAGMA. 

Overall, PRAGMA bridges the gap between correctness-oriented LLM generation and true performance-oriented optimization, delivering faster and more reliable kernel code across diverse workloads.

\subsection{Case Studies}
\label{sec:evaluation:case}
% 多轮性能迭代变化曲线

For better understanding of PRAGMA, we provide detailed case studies on \textit{max reduction over a dimension} task and \textit{conv standard 1D} task in KernelBench.

\subsubsection{Max reduction over a dimension}
\begin{figure}[t]
	\centering
	\includegraphics[width=\linewidth]{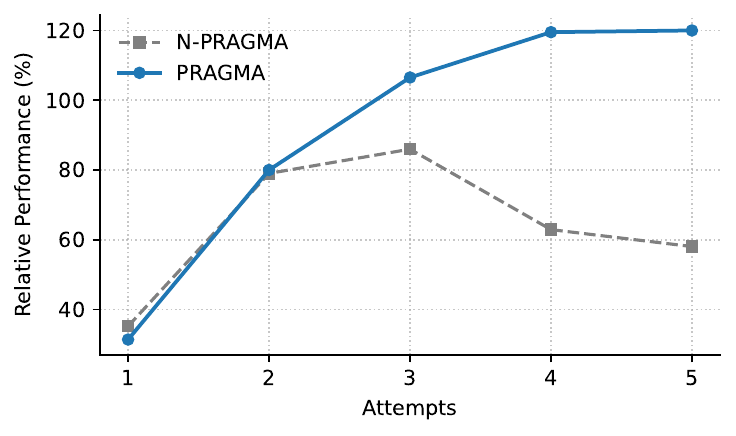}
	\caption{Performance changes of PRAGMA and N-PRAGMA on KernelBench \texttt{Max reduction over a dimension} task over five consecutive attempts.}
	\label{fig:case_reduction}
\end{figure}
We further conduct a detailed analysis on KernelBench \texttt{Max reduction over a dimension} task to illustrate the optimization iteration of PRAGMA.
Figure~\ref{fig:case_reduction} demonstrates the relative performance change curves over Torch across five consecutive attempts on GPU backend.

PRAGMA achieves steady improvement from 31.41\% to 120\%.
In contrast, N-PRAGMA peaks at 85.97\% in the third iteration but later drops to 58.06\%.
The key difference lies in the feedback mechanism.
N-PRAGMA only returns execution time after each iteration, providing coarse-grained guidance.
PRAGMA, in contrast, integrates fine-grained performance metrics such as memory throughput, arithmetic intensity, and occupancy.
These metrics enable precise identification of bottlenecks and direct adjustment of reduction strategies. 
% Meanwhile, PRAGMA preserves historical best-performing code and its corresponding profiling data.
Crucially, the retention of historical best code and its profiling record allows the PRAGMA to judge whether each modification 
yields positive or negative performance impact, enabling stable, performance-aware evolution rather than blind trial-and-error.
% This allows the system to assess whether each new modification improves or degrades performance, enabling the LLM to learn directional feedback.

\subsubsection{Conv Standard 1D}
While most convolution kernels cannot benefit from PRAGMA, a few of them on the CPU backend exhibit significant performance gains, demonstrating the effectiveness of the performance-guided reasoning of PRAGMA.
To better illustrate how PRAGMA leverages performance feedback for iterative optimization,
we present a reasoning trace of the Conductor Agent during two consecutive iterations on the 
KernelBench \texttt{Conv Standard 1D} task on CPU backend.
Each iteration consists of performance comparison, diagnosis, and targeted guidance for code refinement.

As shown in Listing~\ref{lst:conductor_reasoning}, the Conductor Agent analyzes the fine-grained performance counters and provides optimization suggestions.
In the first iteration, the Conductor detects a performance regression despite a higher IPC as shown in Line~14, correctly identifying 
that conditional checks in the vectorized loop degraded efficiency (Line~16).
In the next iteration, after targeted refinement, the Conductor confirms improved performance based on detailed profiling data at Line~31.
It recognizes that the introduction of fused multiply-add (FMA) instructions (Line~33) and reduced instruction count (Line~32) lead to a 
4.71$\times$ speedup over the Torch baseline.
% 补充reasoning的内容展示, aikg_kernelbench_67_conv_standard_1D
\begin{figure}
\begin{lstlisting}[language={}, caption={Excerpt of Conductor Agent reasoning trace in two consecutive iterations.}, label={lst:conductor_reasoning}]
Round 1: Performance regression detected
Historical best:
  Speedup = 2.93x
  IPC = 1.37
  Frontend_bound = 5.10%
  Backend_bound  = 61.76%
Current code:
  Speedup = 2.54x
  IPC = 1.49
  Frontend_bound = 7.68%
  Backend_bound  = 57.24%
Analysis:
  - Speedup dropped from 2.93x to 2.54x
  - Higher IPC but worse performance due to increased branch conditions
  - High backend bound maybe due to memory or core bound
  - Detected inefficient vectorization and irregular memory access
Action:
  Suggest loop refactor for better vectorization and removal of conditional checks in SIMD path

Round 2: Performance improvement recognized
Historical best:
  Speedup = 2.93x
  IPC = 1.37
  Instructions = 945K
Current code:
  Speedup = 4.71x
  IPC = 0.52
  Instructions = 136K
  Cycles = 242K
Analysis:
  - Significant speedup improvement (+60%)
  - Lower IPC but fewer total instructions and cycles
  - Effective FMA vectorization and better data reuse
Action:
  - Mark current code as new historical best
  - Optimize memory access via loop reordering or cache blocking
\end{lstlisting}
\end{figure}

\section{Conclusion}
\label{sec:conclusion}
In this work, we present PRAGMA, a profile-guided multi-agent system for high-performance kernel generation.
PRAGMA bridges the gap between LLM-based code generation and traditional performance-guided optimization by forming a closed feedback loop between profiling, reasoning, and kernel generation. Through multi-backend profiling on both CPU and GPU platforms, PRAGMA transforms low-level performance metrics into interpretable optimization hints and continuously refines kernel implementations.
Our experiments on KernelBench demonstrate that PRAGMA significantly outperforms baseline LLM agents in both success rate and performance, achieving up to 10.95$\times$ speedup while providing transparent and architecture-aware performance insights.
% In particular, the system effectively identifies backend bottlenecks and adapts its strategies to hardware-specific behaviors, improving complex convolution kernels on CPU backends substantially.
We believe PRAGMA represents an important step toward autonomous and high-performance kernel generation.

\bibliographystyle{IEEEtran}
\bibliography{ref}

@article{liu2023your,
  title={Is your code generated by chatgpt really correct? rigorous evaluation of large language models for code generation},
  author={Liu, Jiawei and Xia, Chunqiu Steven and Wang, Yuyao and Zhang, Lingming},
  journal={Advances in Neural Information Processing Systems},
  volume={36},
  pages={21558--21572},
  year={2023}
}

@inproceedings{nguyen2022empirical,
  title={An empirical evaluation of GitHub copilot's code suggestions},
  author={Nguyen, Nhan and Nadi, Sarah},
  booktitle={Proceedings of the 19th International Conference on Mining Software Repositories},
  pages={1--5},
  year={2022}
}

@inproceedings{hong2023metagpt,
  title={MetaGPT: Meta programming for a multi-agent collaborative framework},
  author={Hong, Sirui and Zhuge, Mingchen and Chen, Jonathan and Zheng, Xiawu and Cheng, Yuheng and Wang, Jinlin and Zhang, Ceyao and Wang, Zili and Yau, Steven Ka Shing and Lin, Zijuan and others},
  booktitle={The Twelfth International Conference on Learning Representations},
  year={2023}
}

@inproceedings{wu2024autogen,
  title={Autogen: Enabling next-gen LLM applications via multi-agent conversations},
  author={Wu, Qingyun and Bansal, Gagan and Zhang, Jieyu and Wu, Yiran and Li, Beibin and Zhu, Erkang and Jiang, Li and Zhang, Xiaoyun and Zhang, Shaokun and Liu, Jiale and others},
  booktitle={First Conference on Language Modeling},
  year={2024}
}

@article{qian2023chatdev,
  title={Chatdev: Communicative agents for software development},
  author={Qian, Chen and Liu, Wei and Liu, Hongzhang and Chen, Nuo and Dang, Yufan and Li, Jiahao and Yang, Cheng and Chen, Weize and Su, Yusheng and Cong, Xin and others},
  journal={arXiv preprint arXiv:2307.07924},
  year={2023}
}

@article{ouyang2025kernelbench,
  title={Kernelbench: Can llms write efficient gpu kernels?},
  author={Ouyang, Anne and Guo, Simon and Arora, Simran and Zhang, Alex L and Hu, William and R{\'e}, Christopher and Mirhoseini, Azalia},
  journal={arXiv preprint arXiv:2502.10517},
  year={2025}
}

@article{wen2025multikernelbench,
  title={MultiKernelBench: A Multi-Platform Benchmark for Kernel Generation},
  author={Wen, Zhongzhen and Zhang, Yinghui and Li, Zhong and Liu, Zhongxin and Xie, Linna and Zhang, Tian},
  journal={arXiv e-prints},
  pages={arXiv--2507},
  year={2025}
}

@article{rahman2025marco,
  title={Marco: A multi-agent system for optimizing hpc code generation using large language models},
  author={Rahman, Asif and Cvetkovic, Veljko and Reece, Kathleen and Walters, Aidan and Hassan, Yasir and Tummeti, Aneesh and Torres, Bryan and Cooney, Denise and Ellis, Margaret and Nikolopoulos, Dimitrios S},
  journal={arXiv preprint arXiv:2505.03906},
  year={2025}
}

@article{wei2025astra,
  title={Astra: A multi-agent system for gpu kernel performance optimization},
  author={Wei, Anjiang and Sun, Tianran and Seenichamy, Yogesh and Song, Hang and Ouyang, Anne and Mirhoseini, Azalia and Wang, Ke and Aiken, Alex},
  journal={arXiv preprint arXiv:2509.07506},
  year={2025}
}

@article{tschand2025swizzleperf,
  title={SwizzlePerf: Hardware-Aware LLMs for GPU Kernel Performance Optimization},
  author={Tschand, Arya and Awad, Muhammad and Swann, Ryan and Ramakrishnan, Kesavan and Ma, Jeffrey and Lowery, Keith and Dasika, Ganesh and Reddi, Vijay Janapa},
  journal={arXiv preprint arXiv:2508.20258},
  year={2025}
}

@article{guo2025deepseek,
  title={Deepseek-r1: Incentivizing reasoning capability in llms via reinforcement learning},
  author={Guo, Daya and Yang, Dejian and Zhang, Haowei and Song, Junxiao and Zhang, Ruoyu and Xu, Runxin and Zhu, Qihao and Ma, Shirong and Wang, Peiyi and Bi, Xiao and others},
  journal={arXiv preprint arXiv:2501.12948},
  year={2025}
}

@inproceedings{tillet2019triton,
  title={Triton: an intermediate language and compiler for tiled neural network computations},
  author={Tillet, Philippe and Kung, Hsiang-Tsung and Cox, David},
  booktitle={Proceedings of the 3rd ACM SIGPLAN International Workshop on Machine Learning and Programming Languages},
  pages={10--19},
  year={2019}
}

@article{tay2022efficient,
author = {Tay, Yi and Dehghani, Mostafa and Bahri, Dara and Metzler, Donald},
title = {Efficient Transformers: A Survey},
year = {2022},
issue_date = {June 2023},
publisher = {Association for Computing Machinery},
address = {New York, NY, USA},
volume = {55},
number = {6},
issn = {0360-0300},
url = {https://doi.org/10.1145/3530811},
doi = {10.1145/3530811},
journal = {ACM Comput. Surv.},
month = dec,
articleno = {109},
numpages = {28},
}

@inproceedings{smith202411,
  title={11.1 AMD InstinctTM MI300 series modular chiplet package--HPC and AI accelerator for exa-class systems},
  author={Smith, Alan and Chapman, Eric and Patel, Chintan and Swaminathan, Raja and Wuu, John and Huang, Tyrone and Jung, Wonjun and Kaganov, Alexander and McIntyre, Hugh and Mangaser, Ramon},
  booktitle={2024 IEEE International Solid-State Circuits Conference (ISSCC)},
  volume={67},
  pages={490--492},
  year={2024},
  organization={IEEE}
}

@article{dao2022flashattention,
  title={Flashattention: Fast and memory-efficient exact attention with io-awareness},
  author={Dao, Tri and Fu, Dan and Ermon, Stefano and Rudra, Atri and R{\'e}, Christopher},
  journal={Advances in neural information processing systems},
  volume={35},
  pages={16344--16359},
  year={2022}
}

@article{dao2023flashattention,
  title={Flashattention-2: Faster attention with better parallelism and work partitioning},
  author={Dao, Tri},
  journal={arXiv preprint arXiv:2307.08691},
  year={2023}
}

@article{shah2024flashattention,
  title={Flashattention-3: Fast and accurate attention with asynchrony and low-precision},
  author={Shah, Jay and Bikshandi, Ganesh and Zhang, Ying and Thakkar, Vijay and Ramani, Pradeep and Dao, Tri},
  journal={Advances in Neural Information Processing Systems},
  volume={37},
  pages={68658--68685},
  year={2024}
}

@inproceedings{zhou2025qimeng,
  title={QiMeng-GEMM: Automatically Generating High-Performance Matrix Multiplication Code by Exploiting Large Language Models},
  author={Zhou, Qirui and Wen, Yuanbo and Chen, Ruizhi and Gao, Ke and Xiong, Weiqiang and Li, Ling and Guo, Qi and Wu, Yanjun and Chen, Yunji},
  booktitle={Proceedings of the AAAI Conference on Artificial Intelligence},
  volume={39},
  number={21},
  pages={22982--22990},
  year={2025}
}

@article{zhang2025qimeng,
  title={Qimeng-tensorop: Automatically generating high-performance tensor operators with hardware primitives},
  author={Zhang, Xuzhi and Peng, Shaohui and Zhou, Qirui and Wen, Yuanbo and Guo, Qi and Chen, Ruizhi and Zhu, Xinguo and Xiong, Weiqiang and Chen, Haixin and Ma, Congying and others},
  journal={arXiv preprint arXiv:2505.06302},
  year={2025}
}

@article{zhou2025qimenga,
  title={QiMeng-Attention: SOTA Attention Operator is generated by SOTA Attention Algorithm},
  author={Zhou, Qirui and Peng, Shaohui and Xiong, Weiqiang and Chen, Haixin and Wen, Yuanbo and Li, Haochen and Li, Ling and Guo, Qi and Zhao, Yongwei and Gao, Ke and others},
  journal={arXiv preprint arXiv:2506.12355},
  year={2025}
}

@misc{aikg2025git,
  title={AI-driven Kernel Generator (AIKG)},
  url={https://github.com/mindspore-ai/akg/tree/br_aikg/aikg},
  year={2025}
}

@inproceedings{yasin2014top,
  title={A top-down method for performance analysis and counters architecture},
  author={Yasin, Ahmad},
  booktitle={2014 IEEE International Symposium on Performance Analysis of Systems and Software (ISPASS)},
  pages={35--44},
  year={2014},
  organization={IEEE}
}

\end{document}